**Wavelength Control of Perovskite Metasurface Lasing via Electrical Microheaters**

*Tim Meiler, Yutao Wang, Saurabh Srivastava, Giorgio Adamo, Ramón Paniagua-Dominguez\*, Arseniy Kuznetsov\*, Cesare Soci\**

T. C. Meiler, Y. Wang, G. Adamo, C. Soci

Division of Physics and Applied Physics School of Physical and Mathematical Sciences

Nanyang Technological University

21 Nanyang Link, Singapore 637371, Singapore

E-mail: csoci@ntu.edu.sg

Y. Wang

Interdisciplinary Graduate School, Energy Research Institute @NTU (ERI@N)

Nanyang Technological University

50 Nanyang Drive, Singapore 637553

T. C. Meiler, S. Srivastava, R. Paniagua-Dominguez, A. Kuznetsov

Institute of Materials Research and Engineering

A*STAR (Agency for Science Technology and Research)

2 Fusionopolis Way, #08-03, Innovis, Singapore 138634, Singapore

E-mail: ramon_paniagua@imre.a-star.edu.sg, arseniy_kuznetsov@imre.a-star.edu.sg

Perovskites have recently brought significant advances to active nanophotonics, offering a unique combination of gain and phase-change properties for tunable light-emitting devices. However, current wavelength-tunable devices often rely on tuning mechanisms or device architectures that lead to slow modulation or bulky setups. In this study, we overcome limitations on speed and size by demonstrating a compact tunable microlaser embedding electrical microheaters beneath a perovskite metasurface. This architecture allows to efficiently deliver heat and rapidly modulate the phase transition. Our device leverages the optical gain and crystallographic phase tuning of the perovskite, and a high-quality factor cavity design based on bound states in the continuum. With it, we demonstrate reversible laser wavelength switching between 763 nm and 783 nm within 13 ms at 2.3 V. This work unlocks the potential of perovskite metasurfaces for electrically tunable light sources and introduces a flexible platform which can be easily extended to the dynamic control of polarization or directionality for optical communication, sensing and spectroscopy.

# 1. Introduction

Lasers had a transformative impact on multiple industries such as communications, medicine and manufacturing by enabling high-speed data transmission, precise control over high power light sources and advanced measurement techniques[1]. A key factor for many applications is matching the laser wavelength dynamically to their specific requirements, e.g. to a material absorption band or to a specific wavelength in a multi-channel data processing system. Current tuning methods in semiconductor lasers with microelectromechanical systems or external cavities offer benefits such as room temperature operation but also present challenges for miniaturization due to complex fabrication. These issues require integrating dynamic tunability into both the laser cavity and gain material for compact, efficient and multi-purpose operation.

Compact designs of laser microcavities have received a significant boost in recent years thanks to the use of metasurfaces[2,3], flat arrays of sub-wavelength resonators, which allow to tightly confine light at the nanoscale and precisely control the cavity modes and resonances via fine tuning of the metasurface geometrical parameters. A recent concept, which has drawn strong interest for its promise to, theoretically, deliver perfect field confinement and infinite quality factors, is that of bound states in the continuum (BICs), electromagnetic modes which lie inside the radiation continuum, but do not leak to the far-field[4]. Metasurfaces have demonstrated to be ideal platforms to engineer BICs with tailored properties to achieve desired functionalities such as, for example, lasers with low thresholds[5–7], high non-linearities[8–10] and strong light-matter interaction[11].

Despite the excellent performance of microlasers based on BICs metasurfaces, most of the demonstrations are static, with laser characteristics that remain fixed once fabricated[5,6]. Active metasurfaces aim to incorporate dynamic control over their optical properties[12–14]. Typical methods to achieve this include the use of chemical modifications[15,16], liquid crystals[17,18], external temperature controllers[19] or illumination[20], which are effective but can be challenging to integrate on a micro-scale and to operate at high modulation speeds.

In this regard, electro-thermal tuning through microheaters presents a promising approach, offering precision, speed, and ease of integration for diverse systems[13,21–23]. It is widely adopted in the photonic industry to tune the response of integrated optical components[24–26]. However, common materials only exhibit small changes of optical properties upon thermo-optic tuning, requiring intense heating at the cost of power efficiency and speed. Phase change materials can strongly modulate their refractive index when they transition between phases, but the most prominent materials, vanadium oxide and chalcogenides[27], lack gain for inducing lasing themselves.

In this regard, perovskites have emerged as a novel material platform for developing next-generation optoelectronic devices due to their remarkable optical and electronic properties resulting in both high gain and phase-change behaviour[28–30]. Thus, besides the outstanding progress in low cost, highly efficient solar cells [31–33] and photodetectors [34,35], perovskite lasers have attracted significant attention to realize compact and efficient light sources[36–42] with potential tunability, opening venues for biological imaging[43,44], optical communication, sensing[45,46], advanced display technologies[47,48] and on-chip light sources[49–51].

In this study, we leverage the unique properties of perovskites, i.e. the concurrent presence of optical gain and phase-change behaviour, to demonstrate fast and repeatable wavelength tunability in the visible range in a perovskite metasurface laser, using electrical microheaters. We also investigate the effects of temperature modulation on the laser emission properties, focusing on critical metrics such as wavelength tuning range, stability, and response speed. By integrating metasurfaces and microheater technologies, this work establishes a novel platform for highly controllable and efficient tuneable laser sources on the microscale.

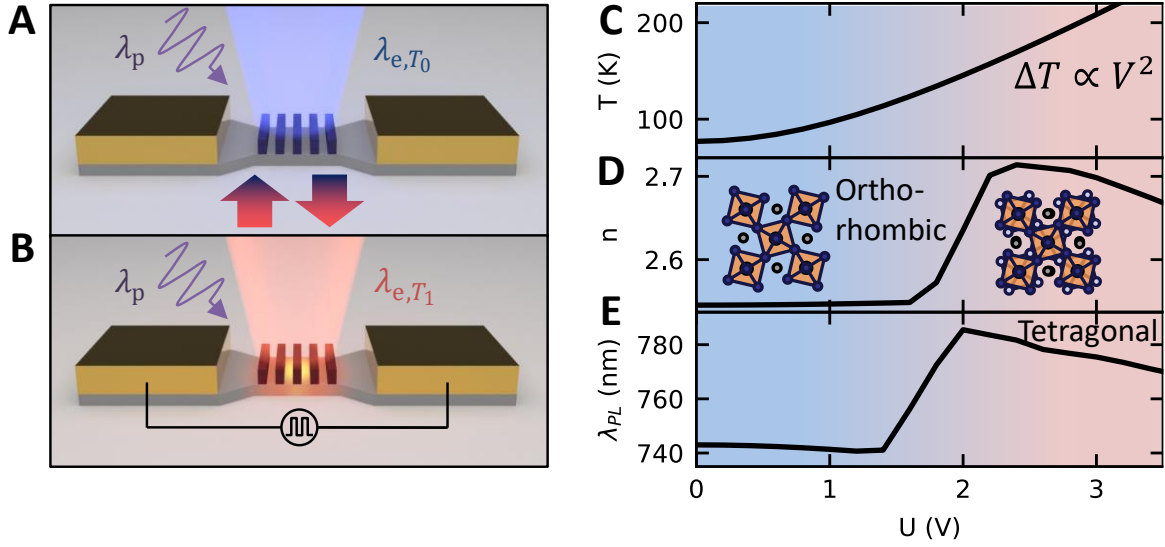

*Figure 1: Electrically actuated laser wavelength switching by microheaters. **A** The cold perovskite metasurface lases at wavelength $\lambda_{e,T_0}$. **B** An electric bias locally heats the microheater below the metasurface above the phase transition temperature causing the laser emission wavelength to redshift to $\lambda_{e,T_1}$. **C** Stationary heat transfer simulations of the microheater temperature T for applied voltage bias U. **D** The refractive index $n(T)$ at 780 nm increases upon phase transition of the methylammonium lead iodide (MAPbI$_3$) perovskite from the tetragonal to orthorhombic phase. **E** The dominating photoluminescence emission wavelength $\lambda_{PL}$ also shifts to the red upon electrothermal heating.*

## 2. Results and Discussion

### 2.1. Wavelength Tunable Perovskite Metasurface Laser using Electrical Microheaters: Concept

Figs. 1**A, B** show a schematic illustration of the principle of operation of our proposed device. An integrated indium tin oxide (ITO) microheater is placed below a methylammonium lead iodide (MAPbI$_3$) perovskite metasurface to locally modify its temperature upon applying electrical bias. Due to the phase change behavior of the perovskite, this enables a simultaneous tuning of its photoluminescence (between two emission wavelengths, $\lambda_{e,T_0}$ and $\lambda_{e,T_1}$) and refractive index, as shown in Figures 1**C-E**. The combination of large optical gain and a high-quality factor BIC cavity in the metasurface design, both tunable with temperature, enables a switchable laser upon optical pumping at wavelength $\lambda_p$ (Figure 1**A**).

The dependence of the MAPbI$_3$ perovskite film temperature on the electrical bias $U$, due to Joule heating of an ohmic resistor $R = U/I$ with power $P$, is quadratic: $\Delta T \propto P = UI = U^2/R$ (see Figure 1**C**). The MAPbI$_3$ perovskite film on top of the microheater reaches its phase transition temperature range of 130-160 K at around 2 V DC bias. Because of structural changes from orthorhombic (low temperature) to tetragonal (high temperature) phase, the refractive index $n$ rises from 2.5 to 2.7 (Figure 1**D**), causing a change in the resonance frequency of the photonic modes in the perovskite metasurface. The same phase transition leads to a 40 nm red-shift of the wavelength of the dominant photoluminescence (PL) emission peak, as shown in Figure 1**E**, thus offering a widely tuneable gain for the device[52].

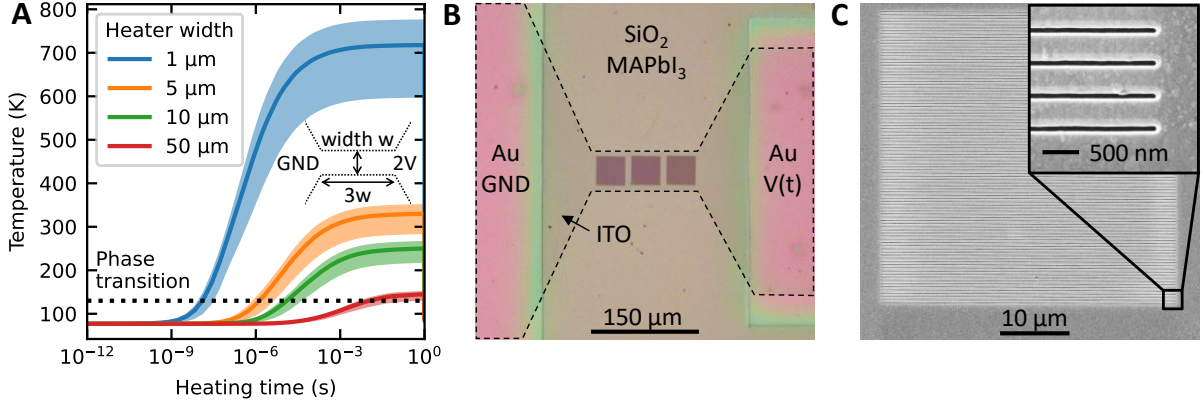

*Figure 2: Microheater and metasurface design. **A** COMSOL heat transfer simulations reveal potential speedup by shrinking microheaters width w. Smaller wires heat up significantly faster at 2 V DC voltage, reaching phase transition temperature below microsecond. **B** Microscope image of multiple metasurfaces on top of a transparent indium tin oxide (ITO) microheater indicated by dashed lines. **C** SEM images of the ion-beam milled nanograting with 430 nm period and 100 nm gap.*

**2.2. Design and Fabrication of Microheater with Metasurface Cavity**

The microheater size is a critical design parameter determining the efficiency and switching speeds of the final device. We performed transient heat transfer simulations to determine its performance as function of the microheater width (see Experimental Methods/Section 4.1): Figure 2A shows the dependence of temperature as function of time, for wire widths ranging from 1 µm to 50 µm after the sudden application of a 2 V DC bias at t=0 s. The smallest microheater (1 µm × 3 µm × 26 nm) considered here would reach the phase transition temperature (~130 K) in 13 ns, consistent with a thermometry study[53]. For lasing, one requires a certain size of the metasurface to guarantee enough gain and large enough quality factor. Consistent with previous works[19], in this work we choose a metasurface size of 40 × 40 µm. To switch it, we use a 50 µm × 150 µm × 26 nm microheater, which reaches the phase transition temperature in 7 ms (Figure 2A). This sets the maximum switching speed in this configuration.

The microscope image in Figure 2B depicts the microlaser device. The microheaters are fabricated from 26 nm indium tin oxide (ITO) film on top of a glass substrate via photolithography (see Experimental Methods/Section 4.2 and Supplementary Figure S1). The transparent ITO wire acts as a localized heat source, which is tapered out towards highly conductive gold contact pads to reduce electrical losses outside the lasing area. A 50 nm layer of HSQ is spin-coated on top of ITO microheaters to passivate the surface before spin-coating a 320 nm thick $MAPbI_3$ film over the whole sample as the active layer for the laser device.

The perovskite metasurface was fabricated by focused ion beam lithography, which mills nanogratings with 430 nm pitch and 330 nm width in the perovskite film on top of the microheater. Figure 2B displays an SEM image of well-defined trenches in the MAPbI$_3$ film (Figure 2C) which form the BIC lasing cavity.

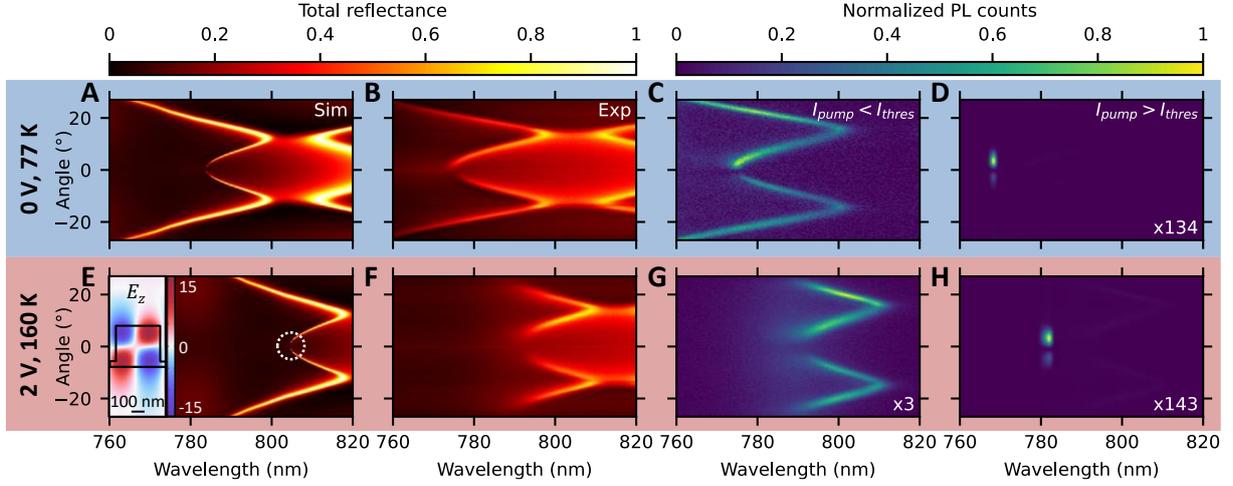

*Figure 3: Angle-resolved reflectance and photoluminescence maps of TE modes in a nanograting. (**A-D**) A cryostat cools the sample to 77 K. (**E-H**) A microheater heats the metasurface above the phase transition temperature under 2 V bias, causing a redshift of the TE modes. **A, E** Simulation and **B, F** experimental results of angle-resolved reflectance. The linewidth of the TE mode narrows towards vertical incidence, indicating a symmetry protected bound state in the continuum (BIC). The inset of **E** shows the near field distribution near the BIC at 810 nm and 2.6° (dashed circle). The electrical field distribution shows a quadrupole resonance along the nanorod, which does not emit into the far-field, achieving high quality factor. **C, G** The photoluminescence (PL) emission below lasing threshold $I_{pump} = 5.5\ \mu J/cm^2$ couples to the modes of the perovskite metasurface in the experiment. **D, H** The BIC lases above the threshold $I_{pump} = 14\ \mu J/cm^2$, but does not emit in vertical direction. **H** The spectral position of the lasing emission redshifts upon the electrical heater bias. The PL maps are normalized with respect to the maximum counts in **C** multiplied with the factor at the bottom right.*

**2.3. Electro-thermal Tuning of Quasi Bound State in the Continuum with Microheaters**

An efficient microlaser design requires good spatial and spectral overlap of the gain medium with the cavity mode. We hence swept several parameters in numerical simulations, e.g. the pitch, width and height of the nanograting, to tune resonances to the emission wavelengths of MAPbI$_3$ (Supplementary Figure S2). The angle-resolved reflectance maps, computed using full-wave numerical simulations (Figure 3A, E), allow to resolve the dispersion of the photonic modes in transverse electric (TE) polarization and compare their evolution as the MAPbI$_3$ film undergoes phase transition (see simulation details in Experimental Section/Methods 4.1). The W-shaped band supports a symmetry protected bound states in the continuum (BIC) at the Γ-point, characterized by the vanishing linewidth under vertical incidence. The BIC is located at around λ=788 nm for temperature of 77 K (Figure 3A) and red shifts by ~20 nm to λ=809 nm at temperature of 160 K (Figure 3E), due to the change in refractive index caused by the MAPbI$_3$ phase transition. The BIC origin can be traced back to an electrical quadrupole resonance oriented along the nanograting, as displayed in the inset of Figure 3E. As such, it does not radiate vertically, as the nodal line of the quadrupole emission is oriented in this direction.

When all unit cells radiate in-phase (corresponding to the Γ-point), radiation into all other directions is suppressed, confining the field inside the nanograting.

To experimentally detect the symmetry protected BIC we use a back-focal plane spectroscopy setup (Supplementary Figure S3) connected to a cryogenic probe station and measure the angle-resolved reflectance maps shown in Figure 3B, F. The measured photonic modes are in good agreement with the simulated ones. Slight asymmetry with regards to the angle might originate from uneven film surface, small tilts of incident light or illuminating the metasurface not directly in the center. A 2 V bias applied to the microheater locally heats the metasurface above the phase transition, causing the BICs to redshift by ~20 nm, from ~774 nm (Figure 3B) to ~794 nm (Figure 3F).

Since the BICs spectrally and spatially overlap with the emission regions of $MAPbI_3$[52], photoluminescence efficiently couples to these optical modes. Angle-resolved photoluminescence maps are presented for unbiased and 2 V bias configurations in Figure 3C and Figure 3G respectively. The metasurface emission closely resembles the reflectance maps, reproducing location and shift of the BIC. As the high-quality factor cavity concentrates the fields inside the gain medium, increasing the fluence of the femtosecond pump laser leads to lasing from the metasurface, when the pump exceeds the threshold of around 13 µJ/cm$^2$, both at the low and high temperature operating regimes. Lasing emission manifests itself as two narrow flat lobes around the normal direction at a wavelength of 768 nm with 1.5 nm full width at half maximum (FWHM) and 6.2° angle between the intensity maxima at 77 K (Figure 3D). They shift to a wavelength of 783 nm with 2.0 nm FWHM and 7.4° angle between them under electrical heating up to 160 K (Figure 3H).

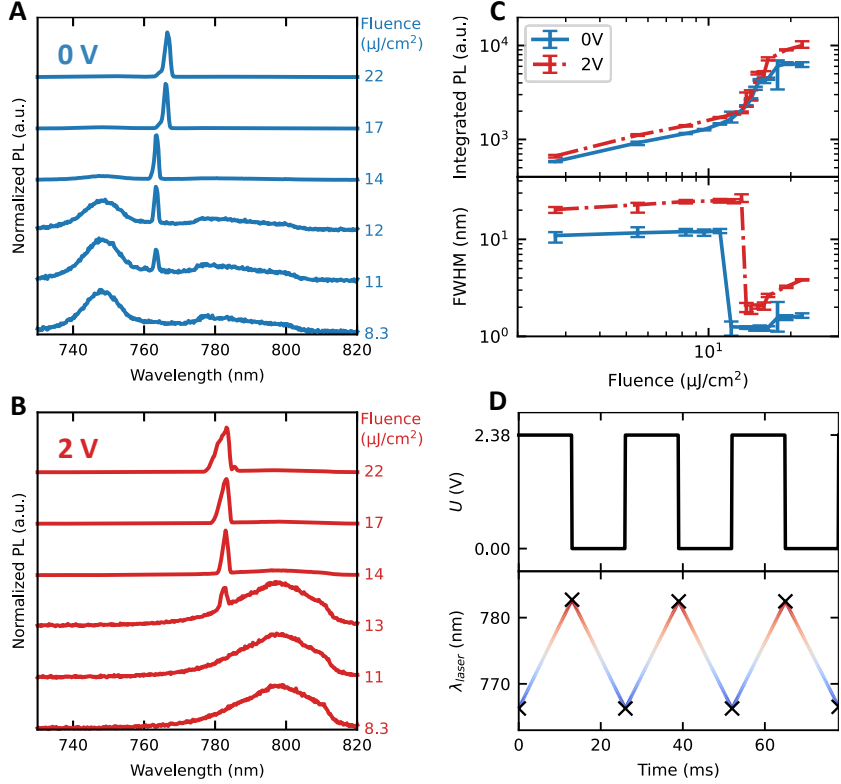

*Figure 4. Lasing action in perovskite metasurface under electrothermal heating. **A**, and **B** Photoluminescence spectra with 0 V (2 V) electrical microheater bias. A narrow lasing peak emerges from the photoluminescence bands, dominating above threshold. **C** Integrated PL intensity (top) and full width at half maximum, FWHM, (bottom) over pump fluence show a threshold behaviour for the BIC laser in both phases. **D** A rectangular voltage bias rapidly switches the BIC wavelength within the spectrometer time resolution of 13 ms.*

### 2.4. Pump Power Dependence of Microlaser

To confirm the lasing nature of the BIC emission reported in Figures 3D and 3H, we further investigate the lasing performance as a function of fluence. PL spectra of Figs. 4A and B, taken for increasing pump laser fluences when the microheater is respectively turned off (0 V) and on (2 V), display narrow lasing peaks at 763 nm and 783 nm arising from broad PL backgrounds in both phases, for pump laser fluences higher than 11 µJ/cm$^2$ and 13 µJ/cm$^2$ respectively. The lasing peaks eventually broaden at high pumping fluences in the saturation regime.

The logarithmic light-out versus light-in curves in the top panel of Figure 4C are calculated from an integral of the PL counts from 700 nm to 820 nm. They grow linearly at low pump fluences and then show a one order of magnitude nonlinear increase above the lasing threshold. Simultaneously, the FWHM (bottom panel in Figure 4C) drops from a 12 nm (24 nm) broad photoluminescence emission to a 1.2 nm (1.8 nm) narrow lasing peak at 0 V (2 V) bias, limited by the spectrometer grating resolution. The quality factor of the lasing cavity, $Q = \lambda_{res}/\mathrm{FWHM}$, is estimated to be higher than 600, in line with previous works[19].

Narrowing of the emission bands, when the pump crosses the threshold, is also visible in the back focal plane photoluminescence images in Supplementary Figure S4, yielding well-defined emission directionality above the threshold. Lastly, lasing is confirmed by coherence fringes appearing in the self-interference measurements (Supplementary Figure S5), which disappear upon increasing the time delay.

**2.5. Dynamic Lasing Switching**

After analyzing the laser characteristics when switching between the two states under constant bias and heating, we tested its performance under dynamic control using voltage pulse trains with 2.38 V, 26 ms period and 50 % duty cycle, as shown in the top panel in Figure 4D. Slightly higher AC voltages are required than in constant voltage case to reach clear phase transitions with maximal modulation depths. A CCD camera continuously acquires photoluminescence spectra, with exposure times of 13 ms, to track the wavelength shift in time. The lasing peak wavelengths clearly switches between 763 nm and 783 nm, following precisely the 13 ms modulation voltage bias period (bottom panel of Figure 4D). Furthermore, rapid switching enables the investigation of durability of our perovskite microlaser over multiple cycles. The video sequence in Supplementary Movie S1 and Supplementary Figure S6 recorded over 2500 switching cycles for 2 min attests an excellent wavelength stability without indications of degradation. Lastly, we also observe continuous lasing wavelength tunability, as shown in Supplementary Figure S7. As we slowly increase the pulse voltage from 2.3 V to 2.6 V, the peak wavelength shifts continuously shifts from 772 nm to 783 nm for the high temperature phase. The emission strength at intermediate wavelengths is slightly reduced and broadened however, as the gain is lower for intermediate phase transition states[52]. Tuning wavelengths in the orthorhombic phase was not tested as it requires lower temperatures than 77 K (liquid nitrogen), which are currently not available in our setup.

## 3. Conclusion

In summary, we realized a metasurface-based microlaser device, which can dynamically switch the lasing wavelength between 763 nm and 783 nm within 13 ms, using a CMOS compatible electrical bias of 2.38 V. The compact 40 µm × 40 µm metasurface is fabricated in a perovskite 350 nm thin $MAPbI_3$ film and combines optical gain and phase-change properties with a high-quality factor dielectric cavity based on bound state in the continuum. Our monolithic approach, which resorts to localize heating of the metasurface via an integrated microheater, sets a benchmark for the switching performance of perovskite phase-change microlasers, an aspect that was lacking in previous proof-of-principle studies[19], and shows that this platform can potentially operate in the microsecond regime and potentially even faster with appropriate scaling of the metasurface dimensions[54,55]. Practical applications would benefit from utilizing materials with phase transitions closer to room temperature such as $BAPb_2I_4$[56] to replace a cryostat with piezoelectrical coolers. Passivation, encapsulation or polymer coatings[57] are techniques that could be employed to guarantee long-term usability of the perovskite films in ambient conditions. Additionally, multiple microheaters in arrays of pixelated elements could achieve advanced displays or phased arrays for light emitting spatial light modulator devices.

## 4. Experimental Section/Methods

*4.1 Numerical Simulations*: Temperature distribution and evolution graphs were simulated using a 3D finite-element method (COMSOL Multiphysics 6.2). The bottom of a 0.7 mm thick glass substrate is kept constant to 77 K, corresponding to the cryostat cold finger temperature used in the experiments. One of the bond pads is grounded while the other one is biased to 2 V. We model the electrical resistivity of ITO at low temperatures with a weak localization model $\rho(T) = \rho_0 + kT^2$[58], measure a sheet resistance of $R_s = 68\ \Omega$ at room temperature with a 4 probe station and compare the resistance of the microheater at room temperature (400 Ω) with 77 K (369 Ω). Stationary simulations result in final temperatures whereas warmup and cooldown rates were calculated with a time dependent solver.

Angle resolved reflectance maps of a perovskite metasurface were simulated using a 2D finite-element methods (COMSOL Multiphysis 6.2). A unit cell consists of a 320 nm thick slab of $MAPbI_3$ (refractive index is taken from [59]) with a 100 nm thin residual layer and horizontal gap on a fused silica substrate (n=1.5) in air (n=1). Floquet periodic boundary conditions were employed in x-direction. A periodic port excites a plane wave from the top with an electric field vector $E = (0,0,1)$ for TE polarization. A second port absorbs it at the bottom.

*4.2 Photolithography of Microheaters:* We inscribed three masks for markers, wires and bond pads in a laser pattern generator (Heidelberg DWL2000). Latech Scientific Supply provided us with 0.7 mm fused silica covered with a 26 nm thin ITO film (>87% transmittance, 70 Ohm/sq sheet resistance). Markers and bond pads are made from a 200 nm thick gold film on a 50 nm thin chromium adhesion layer. A mask aligner (Suss Micro Tec MA6) illuminates the positive photoresist S1811 (Microposit), followed by 1 min development with MF319 developer (Microposit). The metals are wet-etched with standard etchants from Sigma-Aldrich, whereas ITO is dry-etched for 2 minutes in 45 ccm argon and 5 ccm methane with an inductively coupled plasma etching device (Oxford Plasmalab 100 Cobra).

*4.3 Film Preparation*: Thin films of MAPbI$_3$ perovskite were made from a 1 M precursor solution of MAPbI$_3$ (Greatcell Solar Materials) and PbI$_2$ (TCI) with a molar ratio of 1:1 in 1.2 ml anhydrous dimethylformamide (Merck). A magnetic stir bar mixed the solution at 100 °C in a nitrogen glove box for one hour. Afterwards, the solution is run through a PVDF syringe filter with a pore size of 0.45 µm and is kept on a hotplate at 100 °C. Before film deposition, the microheater glass substrates are sonicated in acetone and isopropyl alcohol for 7 minutes each, dried with nitrogen and treated in an ozone cleaner to increase wettability. An antisolvent method was employed to increase nucleation radius and avoid pinholes. The precursor solution was spin-coated onto the microheater substrates at 4900 RPM for 30 s, while drop-casting toluene 5.5 s after the start. The fresh films were annealed at 100 °C for 15 min.

*4.4 Metasurface Nanofabrication*: The 40 µm × 40 µm nanograting patterns were directly written in the film with a focused ion beam (FIB) in the Helios Nanolab 650, which also packages a scanning electron microscope (SEM). The electron beam accelerating voltage was kept low at a 15 kV and 25 pA current to avoid charging of the film and substrate. The ion beam dosage was reduced to 2 mC/cm$^2$ at 30 kV voltage, 6.1 pA current and 1800× magnification to avoid excessive damage to the gain material.

*4.5 Measurement Setup:* Measurements were performed in a cryogenic probe station under vacuum. A schematic of the custom build back focal plane imaging setup[6] with detailed description is included in Supplementary Figure S4. In short, a white light source or a femtosecond laser (405 nm, 60 fs, 1 kHz) excites the perovskite metasurface under normal incidence. The photoluminescence signal is then collected by a 50x objective, passes through a 4f lens setup and a polarization filter to select TE/TM modes before it is analyzed by a CCD grating spectrometer. Angle-resolved reflectance and photoluminescence maps are taken under constant heating bias from a source measure unit, whereas time-resolved spectra are taken under rectangular pulses produced by a wave generator.

**Supporting Information**

Supporting Information is available from the Wiley Online Library or from the author.

**Acknowledgements**

The authors are grateful for advice from Son Tung Ha and Thi Thu Ha Do on lasing and optical setups. We thank Sourav Adhikary for depositing an HSQ passivation layer. This research was supported by the AME programmatic grant A18A7b0058, MTC programmatic grant M21J9b0085, the Quantum Engineering Programme of the Singapore National Research Foundation (NRF2021-QEP2-01-P01), the Singapore Ministry of Education (MOE-T2EP50222-0015). T.C.M. thanks A*STAR Graduate Academy for funding SINGA scholarship.

**Wavelength Control of Perovskite Metasurface Lasing via Electrical Microheaters**

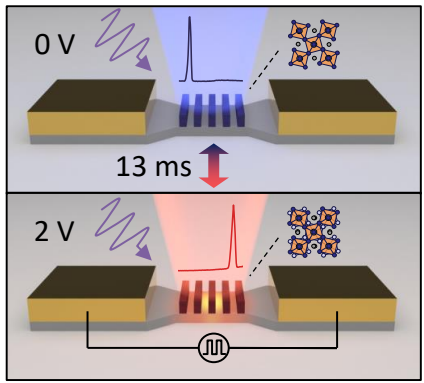

A compact microlaser is efficiently tuned via microheaters beneath a perovskite metasurface. An electrical bias rapidly warms up the gain medium above its phase transition temperature. This device reversibly switches the laser wavelength from 763 nm to 783 nm in 13 ms.

# Supporting Information

## Electrical Tuning of Perovskite Laser via Microheater

*Tim Meiler, Yutao Wang, Saurabh Srivastava, Giorgio Adamo, Ramón Paniagua-Dominguez, Arseniy Kuznetsov\*, Cesare Soci\**

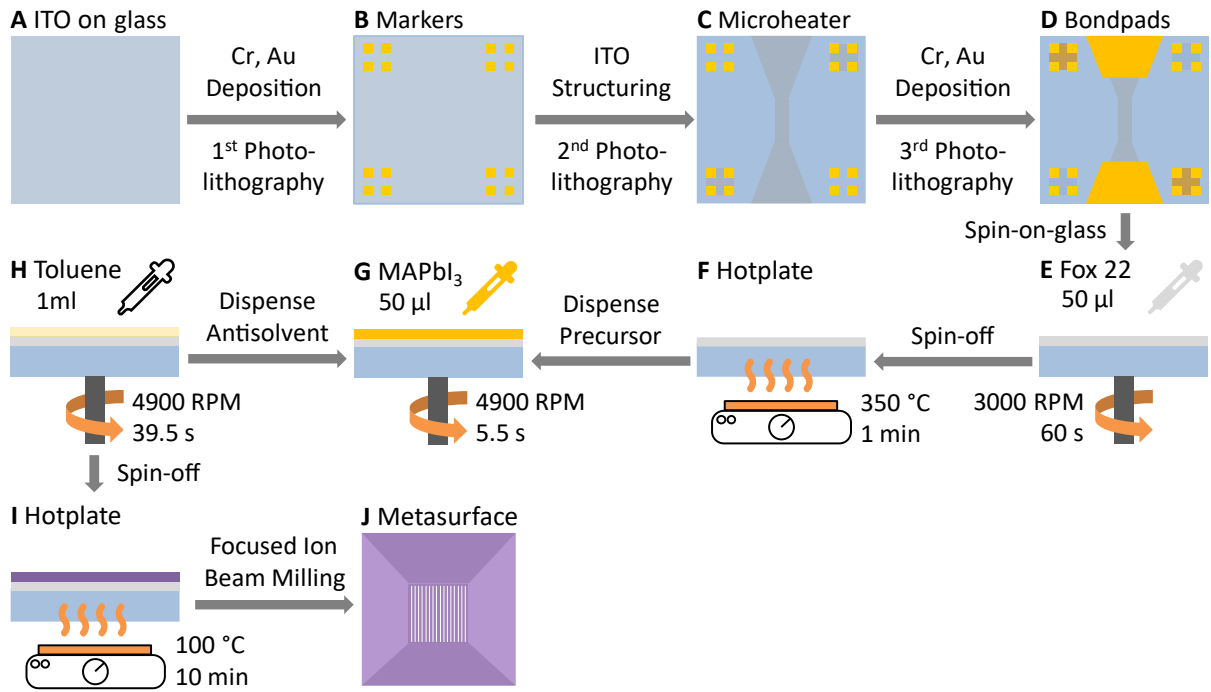

*Figure S1: Schematic of fabrication steps. (**A-D**) Microheaters and bond pads are fabricated via photolithography. (**E-F**) The surface is passivated with spin-on-glass to avoid photoluminescence quenching. (**G-I**) A perovskite film is spin-coated over the whole substrate using an antisolvent technique. (**J**) A focused ion beam mills a metasurface in the perovskite film above the microheater.*

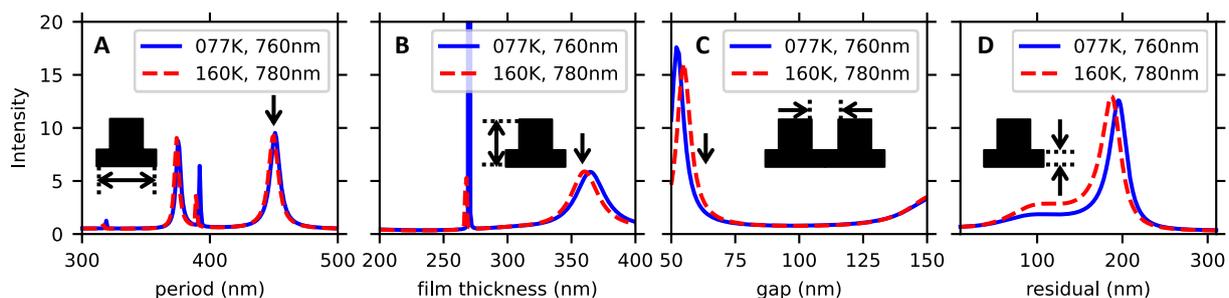

*Figure S2: Metasurface parameter optimization. We sweep over the geometric dimensions of a nanograting to maximize the electric field intensity ($\int \|E\|/\|E_{in}\| dV_{bar}$) inside the gain medium indicated by black arrows over the optima. Simulations need to give optimal values at the emission wavelength of 760 nm (780 nm) in both phases at 77 K (160 K). **A** The electrical quadrupole resonance at 430-450 nm period is enhanced over the broadest parameter range. **B** Varying the precursor concentration and spinning speeds, one can alter the film thickness. Our film quality is optimized at 320 nm thickness which coincides with the broad enhancement peak. The narrow peak at 260 nm is beyond the precision of control over the film thickness and roughness (~6 nm) in spin-coating. **C** The optimal gap of 55 nm is close to the limit of the focused ion beam milling. **D** Incomplete milling can leave a residual layer that can enhance the quality factor through guided resonance modes. We estimate a residual layer of ~100 nm in our experiments from matching location of photonic modes in reflectance measurements and simulations. This parameter is challenging to control via the dosage is difficult to assess in cross-cut SEM due to the softness of the perovskite film.*

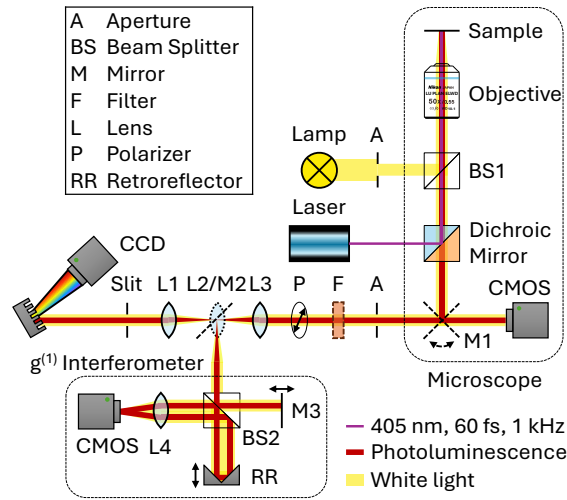

*Figure S3: Back-focal plane imaging setup with g(1) interferometer. A cryostat (LINKAM HFS350EV-PB4) cools the sample down to 77 K and has up to four electrical connections outside to DC power supply (Agilent B2902A) or wave generator (Keithley 3390). When beam splitter BS1 is inserted in an inverted microscope (Nikon Eclipse Ti-U), an integrated white light source illuminates the sample through an infinity corrected objective (Nikon LU Plan 50x 0.55 NA 10.3 WD Epi). A CMOS camera can take real space images if the flip mirror M1 is turned to the right position. Alternatively, a dichroic mirror reflects the pump laser (Coherent Micra, Legend, Evolution and Topas, 405 nm, 60 fs, 1 kHz) to the sample and blocks it from entering the spectrometer (Princeton Instruments Isoplane 320), while photoluminescence can pass through. When the flip mirror M1 is turned to the left position, the light reaches the spectrometer through a 4f setup of the lenses L1, L2 and L3. The CCD camera (Andor iDus DU420A-BEX2-DD) can also image the sample when the diffraction grating (300 lines/mm) is tilted to $0^{th}$ order and Fourier lens L2 is removed. Closing the aperture directly after the microscope to the size of the metasurface can produce cleaner measurements without a line at vertical incidence. The CCD captures the back focal plane image after focusing and moving L2 to the centre of the beam. We then close the slit and set the grating to a central wavelength of 775 nm to measure angle-resolved reflectance or photoluminescence. Polarizer P in the path can select the TE or TM incident polarizations and bandpass filter F can select the lasing wavelengths. This is particularly relevant when determining g(1) correlations in an interferometer, when mirror M2 is inserted and Fourier lens L2 is flipped down. Beam splitter BS2 directs the light beam towards mirror M3 on one arm, whereas retroreflector RR is mounted in the other arm. It flips the image towards the vertical direction, making interference fringes visible on a CMOS camera after collimating lens L4. They disappear after moving one arm of the interferometer by the coherence length.*

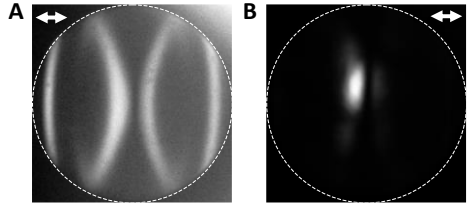

*Figure S4: Back focal plane (BFP) photoluminescence images of the perovskite metasurface. The pictures are taken behind a 780 nm bandpass filter with open spectrometer slit, grating is set to 0 nm and microheater bias to 2 V. **A** If the pump fluence is below the lasing threshold, the photoluminescence of photonic modes of the perovskite metasurface is observed over a large spread of angles. **B** When lasing sets in above the threshold, the emission narrows to two lobes.*

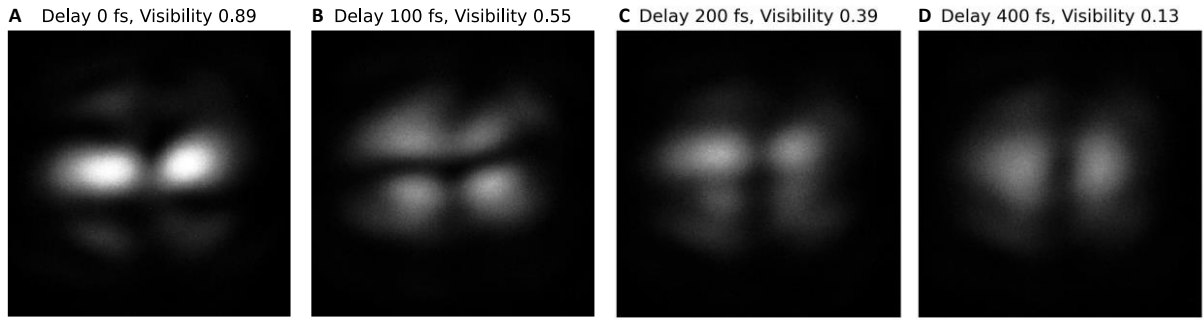

*Figure S5: Self-interference images of the microlaser for temporal coherence measurements. Interference fringes appear when interfering the laser emission with its mirror image. Their visibility decreases upon extending one arm of the interferometer, which introduces a time delay between the interfering signals. The intensity minimum in the vertical direction is due to the cavity design using symmetry protected bound states in the continuum.*

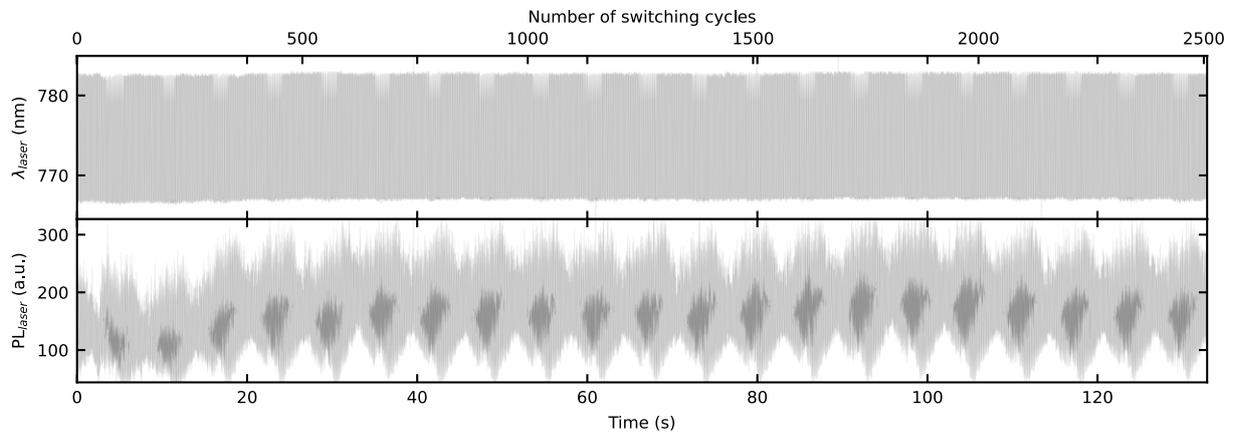

*Figure S6: Stability analysis of the tunable microlaser. Laser wavelengths (top) and photoluminescence intensity (bottom) remain consistent over 2500 switching cycles. The beating pattern in the signal intensity occurs due to slight delays and difference in sampling rates of spectrometer and microheater voltage pulse. Slight pump laser instability $300 \pm 10$ nW and vibrations also cause variations in intensity, but it does not decay overall.*

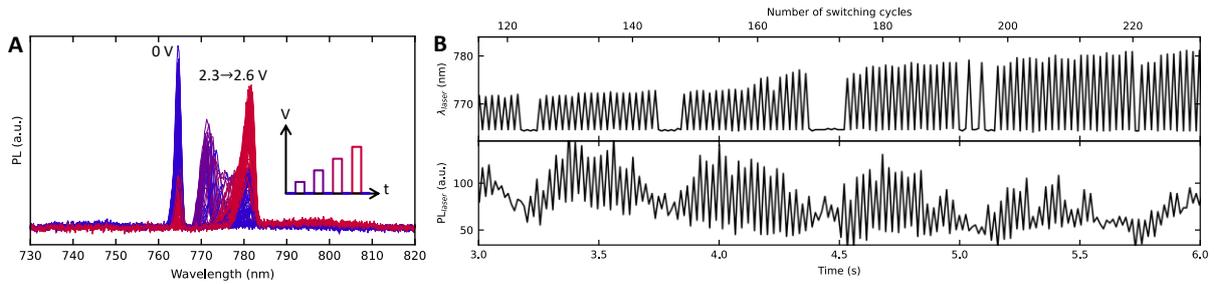

*Figure S7: Continuous wavelength tuning.* **A** Photoluminescence (PL) spectra of perovskite metasurface under ramped heating pulses. The inset illustrates the rectangular voltage pulses with increasing amplitude from 2.3 to 2.6 V with 2 s period (50% duty cycle). **B** Wavelength (top) and amplitude (bottom) of photoluminescence peaks. The peak wavelength sweeps from 770 nm to 783 nm upon increasing the pulse voltage, while it returns to 763 nm during cooldown periods (0 V).

20